\documentclass[a4paper]{article}

\usepackage[nonatbib,final]{metabolisim}

\usepackage[
backend=biber,
style=authoryear,
sorting=nyt,
hyperref=true,
]{biblatex}
\DeclareFieldFormat{citehyperref}{%
	\DeclareFieldAlias{bibhyperref}{noformat}
	\bibhyperref{#1}}
\DeclareFieldFormat{textcitehyperref}{%
	\DeclareFieldAlias{bibhyperref}{noformat}
	\bibhyperref{%
		#1%
		\iffieldundef{postnote}
		{\ifbool{cbx:parens}{\bibcloseparen\global\boolfalse{cbx:parens}}{}}
		{}}}
\savebibmacro{cite}
\savebibmacro{textcite}
\renewbibmacro*{cite}{%
	\printtext[citehyperref]{%
		\restorebibmacro{cite}%
		\usebibmacro{cite}}}
\renewbibmacro*{textcite}{%
	\printtext[textcitehyperref]{%
		\restorebibmacro{textcite}%
		\usebibmacro{textcite}}}

\usepackage[utf8]{inputenc} 
\usepackage[T1]{fontenc}    
\usepackage{amsfonts}       
\usepackage{amssymb}
\usepackage{amsmath}        
\usepackage{graphicx}       
\usepackage{booktabs}       
\usepackage{array}
\usepackage{tabularx}
\usepackage{nicefrac}       
\usepackage{microtype}      
\usepackage{xcolor}         
\usepackage{siunitx}        
\usepackage{hyperref}       
\usepackage{url}            

\graphicspath{{images/}}

\DeclareSIUnit{\kgm}{kg\textsubscript{m}}      
\DeclareSIUnit{\mmHg}{mmHg}

\newcommand{\MetaboliSim}{\textit{MetaboliSim}}

\newcommand{\vo}{\dot{V}\mathrm{O}_2}                 
\newcommand{\vomax}{\dot{V}\mathrm{O}_2\mathrm{max}}  
\newcommand{\vlamax}{\mathrm{VLa_{max}}}              
\newcommand{\maxLass}{\mathrm{maxLa_{ss}}}            

\usepackage{acro}

\DeclareAcroEnding{possessive}{'s}{'s}
\NewAcroCommand\acg{m}{\acropossessive\UseAcroTemplate{first}{#1}}
\NewAcroCommand\acsg{m}{\acropossessive\UseAcroTemplate{short}{#1}}
\NewAcroCommand\aclg{m}{\acropossessive\UseAcroTemplate{long}{#1}}
\NewAcroCommand\acfg{m}{%
	\acrofull
	\acropossessive
	\UseAcroTemplate{first}{#1}%
}
\NewAcroCommand\iacsg{m}{%
	\acroindefinite
	\acropossessive
	\UseAcroTemplate{short}{#1}%
}

\DeclareAcronym{atp}{short=ATP, long=adenosine triphosphate}
\DeclareAcronym{adp}{short=ADP, long=adenosine diphosphate}
\DeclareAcronym{amp}{short=AMP, long=adenosine monophosphate}
\DeclareAcronym{pcr}{short=PCr, long=phosphocreatine}
\DeclareAcronym{gp}{short=GP, long=global phosphate potential}
\DeclareAcronym{ode}{short=ODE, long=ordinary differential equation}
\DeclareAcronym{mlss}{short=MLSS, long=maximal lactate steady state}
\DeclareAcronym{ck}{short=CK, long=creatine kinase}
\DeclareAcronym{ak}{short=AK, long=adenylate kinase}
\DeclareAcronym{pfk}{short=PFK, long=phosphofructokinase}
\DeclareAcronym{pdh}{short=PDH, long=pyruvate dehydrogenase}
\DeclareAcronym{rk4}{short=RK4, long=fourth-order Runge--Kutta}
\DeclareAcronym{rkf45}{short=RKF45, long={Runge--Kutta--Fehlberg (4,5)}}
\DeclareAcronym{amm}{short=AMM, long=active muscle mass}
\DeclareAcronym{rmse}{short=RMSE, long=root mean square error}
\DeclareAcronym{agpl}{short=AGPL, long=GNU Affero General Public License}
\DeclareAcronym{vv}{short=V\&V, long=verification and validation}
\DeclareAcronym{bdf}{short=BDF, long=backward differentiation formula}
\DeclareAcronym{fes}{short=FES, long=Institute for Research and Development of Sport Equipment}

\title{\MetaboliSim: a Python implementation of the Mader model for dynamic
	and steady-state simulation of muscular energy metabolism}

\author{%
	Katharina Dunst\thanks{Corresponding author.} \\
	German Rowing Federation (DRV) \\
	Hannover, Germany \\
	\texttt{katharina.dunst@rudern.de} \\
	\And
	Vincent Scharf \\
	Hochschule Bonn-Rhein-Sieg \\
	Sankt Augustin, Germany \\
	\texttt{vincent.scharf@h-brs.de} \\
	\AND
	Clemens Hesse \\
	German Cycling \\
	Frankfurt, Germany \\
	\texttt{clemens.hesse@germancycling.com} \\
	\And
	Alexander Asteroth \\
	Hochschule Bonn-Rhein-Sieg \\
	Sankt Augustin, Germany \\
	\texttt{alexander.asteroth@h-brs.de}
}

\begin{document}

\maketitle

\begin{abstract}
	The Mader model is the most widely used mathematical framework for muscular energy metabolism in German-language sport science, underpinning lactate diagnostics, maximal lactate steady state (MLSS) estimation and training prescription.
	Despite decades of use, neither its dynamic ODE formulation nor its steady-state equations have been available as open code, leaving results based on the model impossible to reproduce independently.
	We close this gap with \MetaboliSim, an open-source Python implementation of both formulations: a dynamic model that integrates the five-variable ODE system (phosphate potential, $\vo$, muscle and blood lactate, and glycogen) with a fourth-order Runge--Kutta scheme, and a steady-state model that computes MLSS power and the lactate--power relationship in one- and two-compartment variants.
	We verified implementation correctness against published reference values and assessed physiological plausibility across constant-load, step-test, sprint and running protocols.
	The implementation reproduces the published reference output within stated tolerances and remains numerically stable throughout (halving the time step changes blood lactate by less than \qty{0.01}{\milli\mole\per\litre}), with both formulations yielding congruent MLSS estimates.
	Key physiological behaviour ($\vo$ on-kinetics, lactate accumulation, PCr dynamics and the sub/supra-MLSS separation) emerges directly from the model equations without protocol-specific tuning, and a sensitivity analysis shows MLSS power varying approximately linearly with $\vomax$ and nonlinearly with $\vlamax$.
	As the first openly available implementation of the complete Mader model (AGPL-3.0), \MetaboliSim{} lets independent groups reproduce, verify and build on published model-based results.
	Source code: \url{https://codeberg.org/3phos/metabolisim}; Platform: \url{https://metabolisim.org}
\end{abstract}

\acresetall

\section{Introduction}

Quantitative modelling of muscular energy metabolism has a long tradition in exercise physiology.
Since the foundational work of Hill and colleagues on oxygen debt and lactate kinetics, mathematical descriptions of the interplay between oxidative phosphorylation, glycolysis and the phosphocreatine shuttle have evolved from simple compartment models to systems of coupled differential equations that capture the regulatory logic of cellular bioenergetics \parencite{mader2003,maderheck1986}.
Among these, the model developed by Alois Mader and Hermann Heck, referred to here as the Mader model, occupies a central position in German-language sport science and has been applied to lactate diagnostics, performance prediction and training control for several decades \parencite{heck2022}.

The Mader model describes muscular energy metabolism as the interaction of three ATP-supplying pathways---the creatine-kinase equilibrium, anaerobic glycolysis and oxidative phosphorylation---regulated by the concentrations of \ac{adp}, inorganic phosphate and hydrogen ions.
The model exists in two complementary formulations.
The \emph{dynamic formulation} expresses the time evolution of phosphocreatine, oxygen uptake, muscle and blood lactate, and glycogen as a system of coupled \acp{ode}.
The \emph{steady-state formulation} sets all time derivatives to zero and solves the resulting algebraic equations, yielding the relationship between exercise intensity, steady-state lactate concentration and the \acf{mlss}.
Together, the two formulations cover the full range of applications from diagnostics of aerobic and anaerobic capacity to time-course simulation of complex exercise protocols.

Despite its conceptual elegance, no open computational implementation of the Mader model has been available to date.
Mader and Heck computed individual scenarios numerically in the 1980s and 1990s and published the results as figures, but the underlying code was never released \parencite{maderheck1986,mader2003}.
A commercial software tool (INSCYD, Kempten, Germany) implements a variant of the steady-state formulation for applied diagnostics, but its source code is proprietary and the numerical methods are not disclosed.
Consequently, the full model, and in particular the dynamic \ac{ode} formulation, has not been available as inspectable, testable or extensible software.
Published results based on the Mader model remain difficult to reproduce, and the model itself has been less accessible to the international research community than its scientific merit would warrant.

The present work provides the first open-source Python implementation of both Mader model formulations.
It documents the integration scheme and the iterative resolution of the creatine-kinase equilibrium, verifies numerical correctness against published reference values from \textcite{heck2022}, and demonstrates qualitative physiological plausibility through representative simulations for cycling and running.
The contribution is not a new model but a transparent computational realisation of an established physiological framework, supporting independent verification, systematic testing and extension.

\section{Methods}

The methods are organised in two layers.
The first layer---\emph{Conceptual model}, \emph{State-space formulation}, and \emph{Maximum sustainable lactate ($\maxLass$)}---describes the physiological model and is written for readers primarily interested in the sport-science content.
The second layer---\emph{Software architecture}, \emph{Numerical implementation}, and \emph{Parameter handling}---documents the computational realisation and is written for readers primarily interested in the implementation.
The two layers are independent: the physiological description does not require the numerical detail, and the implementation description points back to the relevant equations rather than restating them.
Full rate equations and the default parameter set are given in Appendix~\ref{app:A} and Appendix~\ref{app:B}; for the derivation of the physiological equations the reader is referred to \textcite{mader2003} and \textcite{heck2022}.

\subsection{Notation}

Concentrations and per-mass quantities in this paper use two conventions that are easy to confuse.
To avoid ambiguity, Table~\ref{tab:notation} collects the symbols used throughout. ``Active muscle mass'' denotes the fraction of body mass that is mechanically active during the exercise modelled (\acs{amm}; typically \qty{30}{\percent} of body mass for cycling).
The unit $\mathrm{kg_m}$ always refers to one kilogram of active muscle wet mass.

\begin{table}[t]
	\caption{Notation used throughout the paper.}
	\label{tab:notation}
	\centering
	\small
	\begin{tabularx}{\linewidth}{@{}llX X@{}}
		\toprule
		\textbf{Symbol} & \textbf{Unit} & \textbf{Meaning} & \textbf{Notes} \\
		\midrule
		$\mathrm{kg_m}$ & \unit{\kilo\gram} & Active muscle mass (wet) & Mass unit, not $\mathrm{kg}\cdot\mathrm{m}$ torque \\
		$[\mathrm{X}]_m$ & \unit{\milli\mole\per\litre} & Intramuscular concentration of X & Per L muscle intracellular water \\
		$[\mathrm{X}]_b$ & \unit{\milli\mole\per\litre} & Blood concentration of X & Per L blood \\
		$[\mathrm{X}]_{\mathrm{kgm}}$ & \unit{\milli\mole\per\kgm} & Per active muscle mass amount of X & $= [\mathrm{X}]_m \cdot V_{\mathrm{rel}}$, with $V_{\mathrm{rel}} \approx \qty{0.75}{\litre\per\kgm}$ \\
		$\vo$ & \unit{\milli\litre\per\second\per\kgm} & Oxygen uptake of active muscle & Converted to \unit{\milli\litre\per\minute\per\kgm} or \unit{\milli\litre\per\minute\per\watt} where stated \\
		$\mathrm{GP}$ & \unit{\milli\mole\per\kgm} & Global phosphate potential $\mathrm{ATP}+\mathrm{PCr}$ & Integrated state variable, see Methods \\
		$P$ & \unit{\watt} & External mechanical power & Cycling load on ergometer; running uses \unit{\metre\per\second} \\
		MLSS & --- & Maximal lactate steady state & Highest power with finite steady-state $[\mathrm{La}]_b$ \\
		$\maxLass$ & \unit{\milli\mole\per\litre} & $[\mathrm{La}]_{b,\mathrm{ss}}$ at MLSS in two-compartment model & See Methods: Maximum sustainable lactate \\
		AMM & \unit{\percent} & Active muscle mass as fraction of body mass & Athlete-specific, default \qty{30}{\percent} for cycling \\
		$V_{\mathrm{rel}}$ & \unit{\litre\per\kgm} & Intracellular water volume per $\mathrm{kg_m}$ & Converts $[\mathrm{X}]_m \leftrightarrow [\mathrm{X}]_{\mathrm{kgm}}$ \\
		\bottomrule
	\end{tabularx}
\end{table}

Default values of the kinetic and structural parameters are given in Appendix~\ref{app:B} (Table~\ref{tab:params}).
Subscript conventions: ``ss'' denotes steady state, ``ox'' oxidation, ``res'' resynthesis (gluconeogenesis), ``m'' muscle compartment, ``b'' blood compartment.

\subsection{Conceptual model}

\begin{figure}[t]
	\centering
	\includegraphics[width=\linewidth]{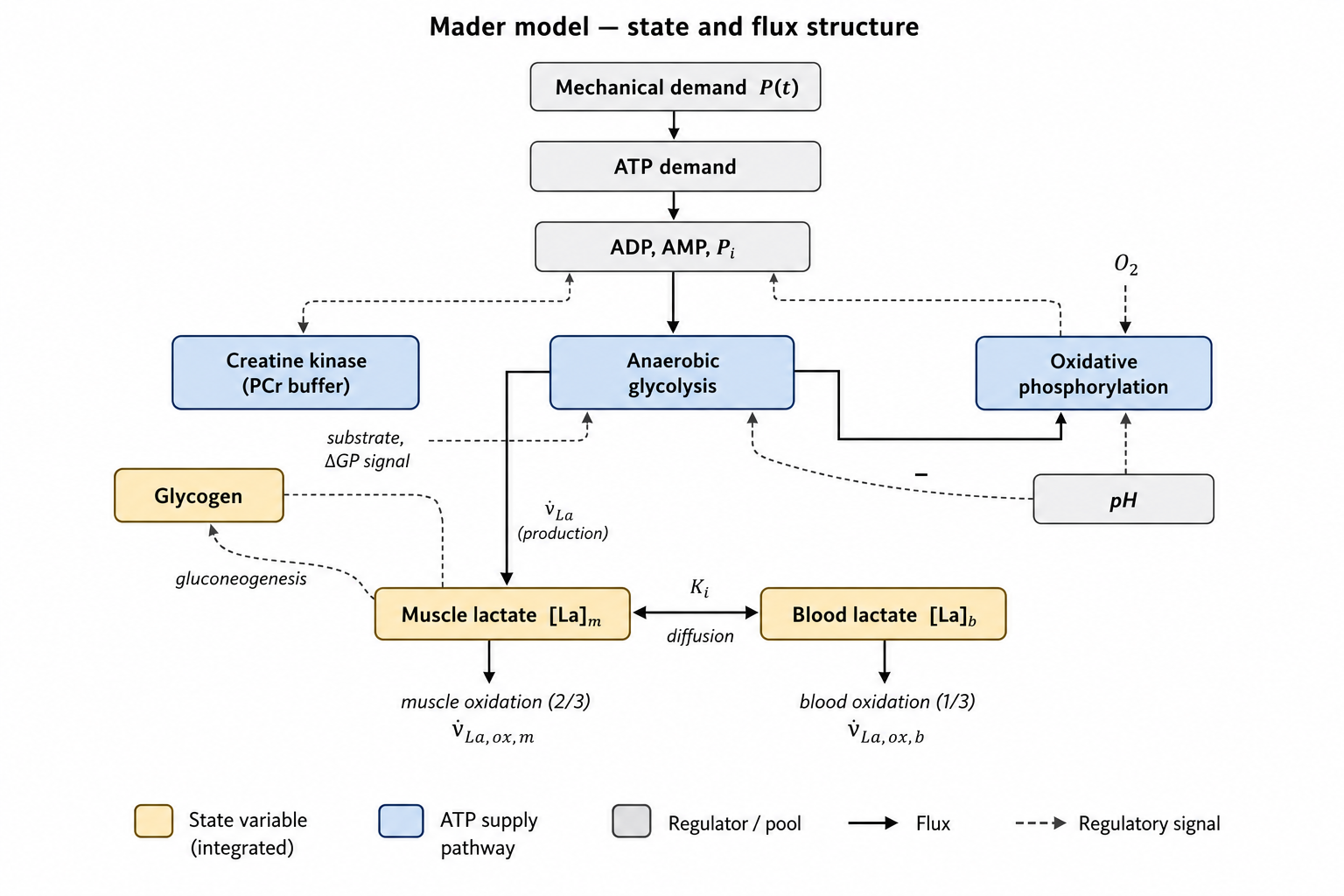}
	\caption{Mader model---state and flux structure.
	Mechanical demand drives ATP consumption, which is met by three parallel supply pathways.
	The $\mathrm{ADP}$, $\mathrm{AMP}$ and $\mathrm{P_i}$ pool functions as the central regulator that couples demand to supply; pH provides negative feedback on glycolysis.
	Lactate is produced by glycolysis, diffuses between muscle and blood compartments, and is removed by oxidation in both compartments.
	Five state variables (yellow) are integrated in the dynamic formulation: $\mathrm{PCr}$ (via the global phosphate potential $\mathrm{GP}=\mathrm{ATP}+\mathrm{PCr}$), $\vo$, $[\mathrm{La}]_m$, $[\mathrm{La}]_b$, and glycogen.}
	\label{fig:structure}
\end{figure}

The Mader model describes the ATP turnover of a working skeletal muscle as the balance of three supply pathways and a mechanical demand.
The pathways are: (i)~the \acf{ck} equilibrium, which buffers ATP concentration through the \acf{pcr} shuttle; (ii)~anaerobic glycolysis, which produces ATP and lactate from glycogen; and (iii)~oxidative phosphorylation, which re-synthesises ATP in the mitochondria using oxygen.
A fourth process, gluconeogenesis, resynthesises glycogen from lactate at an energetic cost.
Lactate is distributed between the muscle and blood compartments by concentration-dependent diffusion and is removed by oxidation in both compartments \parencite{mader2003,heck2022}.
Figure~\ref{fig:structure} summarises the state and flux structure.

The regulatory structure rests on three central metabolite signals.
First, \acf{adp} concentration, derived from the coupled \ac{ck} and \acf{ak} equilibria, activates both oxidative phosphorylation and glycolysis via Michaelis--Menten kinetics with Hill-type cooperativity.
Second, intracellular pH, which decreases with lactate accumulation and phosphate release, inhibits glycolysis through allosteric regulation of \acf{pfk}.
Third, glycogen availability modulates both glycolytic capacity and, to a lesser extent, oxidative capacity.
These regulatory loops create the characteristic nonlinear behaviour of the system: at low intensities, oxidative phosphorylation meets the ATP demand with minimal lactate accumulation; at higher intensities, glycolytic flux increases, lactate accumulates and pH falls, partially inhibiting further glycolysis; above the \ac{mlss}, lactate production exceeds the maximal elimination capacity and no steady state exists.

The dynamic and steady-state formulations describe the same physiology at different levels of temporal resolution.
The steady-state solution represents the fixed point of the dynamic system, and the two-compartment steady-state model uses the identical rate equations as the \ac{ode} system.

\subsection{State-space formulation}

The dynamic model integrates five state variables (Table~\ref{tab:states}): the global phosphate potential $\mathrm{GP}=\mathrm{ATP}+\mathrm{PCr}$, oxygen uptake, muscle and blood lactate concentrations, and muscle glycogen.
The choice of \ac{gp} rather than \ac{pcr} as the integrated variable follows \textcite{heck2022} and avoids an implicit algebraic constraint between ATP and PCr at each step; PCr is recovered iteratively from \ac{gp} after each integration step via the \ac{ck}\,/\,\ac{ak} equilibrium (Appendix~\ref{app:A}, Eqs.~\ref{eq:A1}--\ref{eq:A4}).

\begin{table}[t]
	\caption{State variables of the dynamic model.}
	\label{tab:states}
	\centering
	\small
	\begin{tabular}{@{}lllp{4.6cm}@{}}
		\toprule
		\textbf{Variable} & \textbf{Symbol} & \textbf{Unit} & \textbf{Description} \\
		\midrule
		Global phosphate potential & $\mathrm{GP}$ & \unit{\milli\mole\per\kgm} & $\mathrm{ATP}+\mathrm{PCr}$ \\
		Oxygen uptake & $\vo$ & \unit{\milli\litre\per\second\per\kgm} & Active muscle $\mathrm{O_2}$ consumption \\
		Muscle lactate & $[\mathrm{La}]_m$ & \unit{\milli\mole\per\litre} & Intracellular water \\
		Blood lactate & $[\mathrm{La}]_b$ & \unit{\milli\mole\per\litre} & Blood concentration \\
		Glycogen & $\mathrm{Gly}$ & \unit{\gram\per\kgm} & Muscle glycogen \\
		\bottomrule
	\end{tabular}
\end{table}

In compact form, the dynamic model is a non-autonomous \ac{ode} system
\begin{equation}
	\frac{d\mathbf{x}}{dt} = f\bigl(\mathbf{x}(t), P(t); \boldsymbol{\theta}\bigr),
	\qquad \mathbf{x}(0) = \mathbf{x}_0,
	\tag{M1}\label{eq:M1}
\end{equation}
with state vector $\mathbf{x} = (\mathrm{GP}, \vo, [\mathrm{La}]_m, [\mathrm{La}]_b, \mathrm{Gly})^{\mathsf{T}}$, time-varying mechanical load $P(t)$, and parameter vector $\boldsymbol{\theta}$ (Appendix~\ref{app:B}).
The right-hand side $f$ couples the three ATP supply pathways and the lactate compartments shown in Fig.~\ref{fig:structure}; its components are derived from Michaelis--Menten kinetics with Hill cooperativity for the \ac{adp}-dependent activation of glycolysis and oxidative phosphorylation, pH-dependent inhibition of glycolysis, and concentration-driven diffusion between muscle and blood.
The full functional form is given in Appendix~\ref{app:A} (Eqs.~\ref{eq:A5}--\ref{eq:A16}); the original derivation is in \textcite{mader2003}, with the formulation used here following \textcite{heck2022}.

\subsection{Maximum sustainable lactate (\texorpdfstring{$\maxLass$}{maxLa\_ss})}

A specific use of the steady-state formulation is the identification of the \acf{mlss}---the highest exercise intensity at which a stable blood-lactate equilibrium exists---and the corresponding blood-lactate concentration, here denoted $\maxLass$.
This section makes the underlying aggregated balance explicit, because both the conceptual interpretation of \ac{mlss} and the practical algorithm hinge on it.

\begin{figure}[t]
	\centering
	\includegraphics[width=\linewidth]{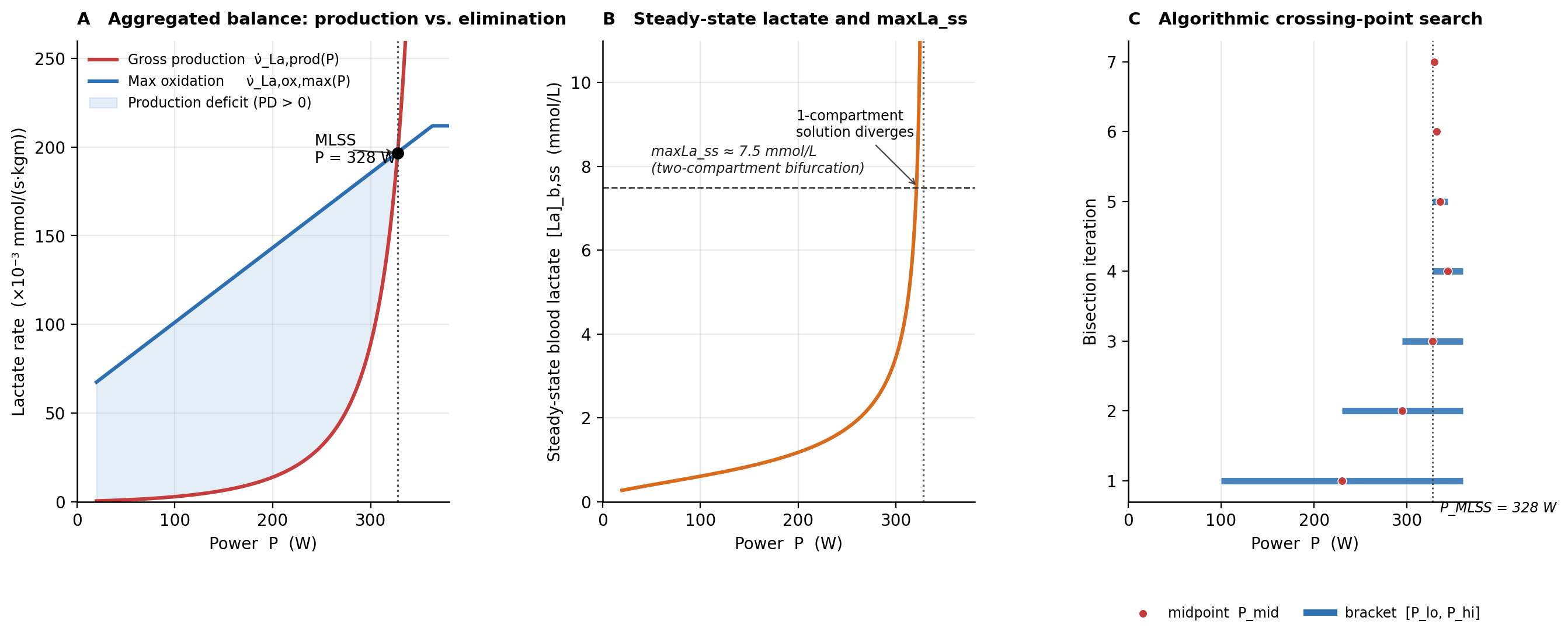}
	\caption{Maximum sustainable lactate.
	(A)~Aggregated balance: gross glycolytic lactate production (red) and maximal lactate oxidation capacity (blue) as functions of power.
	Their intersection defines the \acs{mlss} power; below the \acs{mlss} the production deficit $\mathrm{PD}$ is positive (shaded).
	(B)~Steady-state blood lactate.
	The one-compartment solution diverges at the \acs{mlss}; the two-compartment formulation produces a finite $\maxLass$ at the bifurcation.
	(C)~Bisection algorithm finding the \acs{mlss} power: successive brackets (blue) narrow around the crossing-point midpoint (red).
	Curves illustrative; numerical values from the default parameter set of Appendix~\ref{app:B}.}
	\label{fig:maxlass}
\end{figure}

\paragraph{Aggregated balance.}
At any constant load $P$, the dynamic system reaches a steady state when the gross lactate production from glycolysis equals the maximum lactate elimination via oxidation:
\begin{equation}
	\nu_{\mathrm{La,prod}}(P) = \nu_{\mathrm{La,ox,max}}(P).
	\tag{M2}\label{eq:M2}
\end{equation}
Both sides depend on $P$ through the steady-state oxygen uptake $\vo(P)$ and the resulting \ac{adp}, pH and $[\mathrm{La}]_m$ levels.
The production deficit
\begin{equation}
	\mathrm{PD}(P) = \nu_{\mathrm{La,ox,max}}(P) - \nu_{\mathrm{La,prod}}(P)
	\tag{M3}\label{eq:M3}
\end{equation}
is positive below the \ac{mlss}, zero at the \ac{mlss}, and negative above.
Substituting into the one-compartment lactate balance (Appendix~\ref{app:A}, Eq.~\ref{eq:A17}) yields the steady-state blood lactate
\begin{equation}
	[\mathrm{La}]_{b,\mathrm{ss}}(P) =
	\sqrt{\frac{K_{\mathrm{el}}\,\nu_{\mathrm{La,prod}}(P)}{\mathrm{PD}(P)}}.
	\tag{M4}\label{eq:M4}
\end{equation}
In the one-compartment formulation, $[\mathrm{La}]_{b,\mathrm{ss}}$ diverges as $\mathrm{PD}\to 0$; the \ac{mlss} power is therefore defined as the unique root of $\mathrm{PD}(P)=0$ and found by bisection on Eq.~\eqref{eq:M3}.
The two-compartment formulation (Appendix~\ref{app:A}, Eq.~\ref{eq:A18}) does not exhibit divergence: the muscle and blood lactate equations form a $2\times 2$ nonlinear system whose Jacobian becomes singular at the bifurcation, giving a finite $\maxLass$.
This bifurcation is detected by tracking the largest real part of the Jacobian eigenvalues; the power at which it crosses zero defines the bifurcation point (and therefore the two-compartment \ac{mlss}), and the lactate concentration at that point is $\maxLass$.
Figure~\ref{fig:maxlass} visualises the aggregated balance, the steady-state lactate curve, and the bisection algorithm.

\paragraph{Why this matters.}
Eqs.~\eqref{eq:M2}--\eqref{eq:M4} make explicit that the \ac{mlss} power and $\maxLass$ are not measured quantities in the model but emergent properties of the balance between glycolytic production and oxidative elimination.
Their values are therefore directly traceable to the underlying kinetic constants (Appendix~\ref{app:B})---and any uncertainty in those constants propagates into the predicted \ac{mlss}.
The bisection step is fast and deterministic; in \MetaboliSim{} it forms the core of the steady-state solver and is reused by the parameter-estimation workflow.

\subsection{Software architecture}

\begin{figure}[t]
	\centering
	\includegraphics[width=\linewidth]{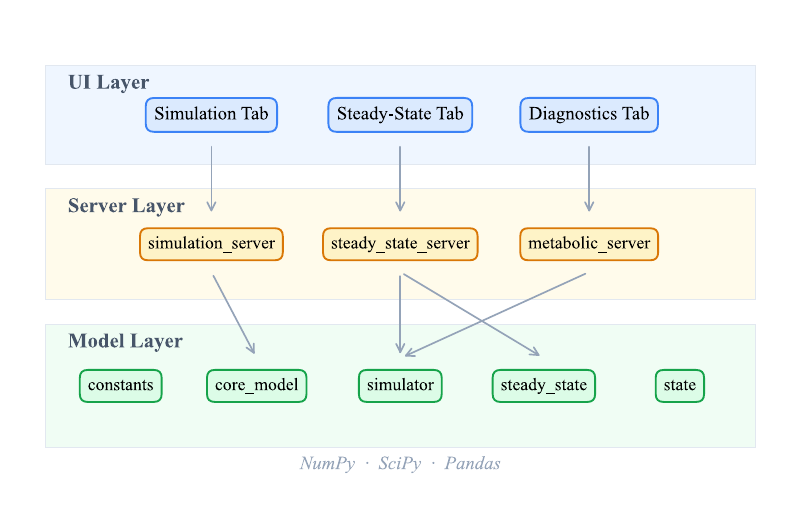}
	\caption{Modular software architecture of \MetaboliSim.
	The model layer (green) contains all physiological equations and numerical methods.
	The server layer (amber) orchestrates reactive computations.
	The UI layer (blue) provides the interactive web interface.
	Dependencies flow downward only; the model layer has no UI dependencies.}
	\label{fig:architecture}
\end{figure}

\MetaboliSim{} is structured around three principles: separation of physiological logic from numerical implementation and from user interface; immutability of parameter sets for reproducibility; and testability of each component in isolation (Fig.~\ref{fig:architecture}).
The model layer contains all physiological equations and numerical methods with no UI dependencies.
It consists of six modules: constants (immutable parameter dataclasses), core model (stateless rate equations), simulator (\ac{ode} integration), steady-state solver, state vector definition and load profiles.
The server layer orchestrates reactive computations for the Shiny web framework.
The UI layer defines the interface as declarative Shiny modules with Plotly visualisations.
The model layer depends only on NumPy, SciPy and Pandas.

\subsection{Numerical implementation}

\paragraph{Integration scheme.}
The \ac{ode} system (Eq.~\eqref{eq:M1}; full form in Appendix~\ref{app:A}, Eqs.~\ref{eq:A12}--\ref{eq:A16}) is integrated with a classical \acf{rk4} method.
At each time step, the computation proceeds as follows: pH is estimated from the previous step's phosphate and lactate concentrations; PCr is recovered from \ac{gp} by Newton iteration on the \ac{ck}\,/\,\ac{ak} equilibrium, yielding \ac{adp}; the metabolic rates (glycolysis, oxidation, lactate exchange, gluconeogenesis) are then computed from the current \ac{adp}, pH and lactate.
Following the strategy described by \textcite{mader2003}, all these rates are evaluated once per step and held constant across the four \ac{rk4} stages (\emph{frozen metabolic coefficients}).
This ensures numerical determinism and avoids oscillatory coupling between fast (\ac{ck}\,/\,\ac{ak} equilibration on the millisecond scale) and slow (lactate diffusion on the minute scale) subsystems.
The time step is adapted to the simulation duration ($\Delta t = \qtyrange{0.05}{1.0}{\second}$).
An adaptive \acf{rkf45} method with embedded error estimation is available as an alternative (default tolerance: \num{e-6}).

\paragraph{Recovery of PCr from GP.}
Because the integrated variable is $\mathrm{GP}=\mathrm{ATP}+\mathrm{PCr}$, PCr must be recovered at each time step by solving the nonlinear equation
\begin{equation}
	f(\mathrm{PCr}) = \mathrm{ATP}(\mathrm{PCr}, \mathrm{pH}) + \mathrm{PCr} - \mathrm{GP} = 0,
	\tag{M5}\label{eq:M5}
\end{equation}
where $\mathrm{ATP}(\mathrm{PCr}, \mathrm{pH})$ is determined by the \ac{ck}\,/\,\ac{ak} equilibrium (Appendix~\ref{app:A}, Eqs.~\ref{eq:A1}--\ref{eq:A4}).
Newton iteration with the analytical derivative $f'(\mathrm{PCr}) = d\mathrm{ATP}/d\mathrm{PCr} + 1$ converges to machine precision in 4--6 iterations.
A bisection fallback guarantees convergence when the Newton step would exceed the feasible interval.

\paragraph{Steady-state solver.}
The one-compartment steady-state model evaluates Eq.~\eqref{eq:M4} directly over a power grid and locates the \ac{mlss} by bisection on Eq.~\eqref{eq:M3}.
When pH feedback is enabled, $\vlamax$ is reduced by the pH inhibition factor and the system is solved by fixed-point iteration.
The two-compartment model uses three nested numerical procedures: a bisection--Newton hybrid for PCr ($d\mathrm{GP}/dt = 0$), a fixed-point loop for muscle lactate, and a bisection for blood lactate.
\ac{mlss} stability is assessed by the eigenvalues of the $2\times 2$ Jacobian of the lactate subsystem as described in the previous section.

\paragraph{Numerical safeguards.}
All state variables are bounded to physiologically meaningful ranges after each step.
Logarithmic arguments are guarded against non-positive values, and the ATP/ADP ratio $Q$ is clamped to prevent overflow in recovery.
The convergence of Newton, the boundedness of state variables and the residual of the steady-state fixed-point iteration are all monitored as part of the automated test suite (see Verification).

\subsection{Parameter handling}

All parameters are defined in a single location with explicit default values, physical units and references to the corresponding equations.
The \texttt{MaderConstants} dataclass serves as the single source of truth.
Athlete-specific parameters (body mass, active muscle mass, $\vomax$, $\vlamax$) are collected in the \texttt{AthleteProfile} dataclass.
The default parameter set is given in Appendix~\ref{app:B} (Table~\ref{tab:params}).
With the exception of the cytosolic \ac{ck} equilibrium constant $M_2$ \parencite{veech1979}, all default parameters are taken from \textcite{mader2003}, \textcite{maderheck1994} or \textcite{heck2022}; none has been independently re-fitted in the present work.

\subsection{Numerical properties and plausibility assessment}

Following the \acf{vv} terminology of \textcite{roache1998}, the present work establishes \emph{numerical properties} of the implementation---that the equations are solved correctly and reproducibly---rather than \emph{validation} against experimental data.
Five procedures are applied: (i)~convergence to resting equilibrium at zero load; (ii)~reproduction of the published reference values from \textcite{heck2022} under standardised conditions (\qty{50}{\watt}, \qty{600}{\second}; PCr: $16.459 \pm 0.2~\unit{\milli\mole\per\kgm}$, $[\mathrm{La}]_b$: $1.096 \pm 0.05~\unit{\milli\mole\per\litre}$); (iii)~exercise-mode equivalence (\qty{3.0}{\metre\per\second} vs \qty{10.8}{\kilo\metre\per\hour}); (iv)~Newton convergence of the \ac{ck}\,/\,\ac{ak} equilibrium at every time step; and (v)~dynamic--steady-state congruence below the \ac{mlss}.
Detailed pass/fail criteria for each procedure are implemented as automated test scripts.

Physiological plausibility is assessed against established qualitative behaviour: mono-exponential $\vo$ on-kinetics, intensity-dependent lactate accumulation, PCr depletion and recovery dynamics, \ac{mlss} within plausible ranges, and the fat-oxidation crossover.
These assessments confirm that the model output is consistent with known exercise physiology but do not constitute a formal validation against experimental data; we return to this distinction in the Limitations section.

\section{Results}

\subsection{Numerical properties}

The \ac{rk4} integrator with frozen coefficients produces stable, non-oscillatory solutions for all tested protocols, including supramaximal sprints, abrupt load transitions and prolonged recovery phases.
The Newton solver converges within 4--6 iterations to residuals below \qty{e-12}{\milli\mole\per\kgm} at every time step without requiring the bisection fallback.
Halving the time step from \qty{0.1}{\second} to \qty{0.05}{\second} changes 5-minute blood lactate by less than \qty{0.01}{\milli\mole\per\litre}, indicating low sensitivity to time-step variation in the tested range.

Under the reference parameter set from \textcite{heck2022}, a \qty{50}{\watt} constant load for \qty{600}{\second} (\qty{75}{\kilo\gram}, \qty{30}{\percent}~\acs{amm}, $\vomax = \qty{50}{\milli\litre\per\minute\per\kilo\gram}$, $\vlamax = \qty{0.5}{\milli\mole\per\litre\per\second}$) produces $\mathrm{PCr} = \qty{16.46}{\milli\mole\per\kgm}$ and $[\mathrm{La}]_b = \qty{1.09}{\milli\mole\per\litre}$, matching the published reference values within specified tolerances.
Exercise-mode equivalence (\qty{3.0}{\metre\per\second} vs \qty{10.8}{\kilo\metre\per\hour}) is confirmed with PCr differences below \qty{0.001}{\milli\mole\per\kgm}.

A \qty{300}{\second} constant-load simulation at $\Delta t = \qty{0.5}{\second}$ completes in approximately \qty{50}{\milli\second} on a standard desktop CPU (single core, Intel-class x86-64, Python~3.11).
The steady-state model computes the full intensity range ($N = 900$ power points from \qtyrange{50}{500}{\watt} in \qty{0.5}{\watt} steps) in approximately \qty{15}{\milli\second} without pH feedback and \qty{80}{\milli\second} with pH feedback.
A parameter sweep across nine $\vomax$ values (\qtyrange{40}{80}{\milli\litre\per\minute\per\kilo\gram} in \qty{5}{\milli\litre\per\minute\per\kilo\gram} steps) requires less than \qty{150}{\milli\second}.
These execution times place the model well within the range required for iterative optimisation, batch parameter fitting and interactive use.

\subsection{Physiological plausibility}

\begin{figure}[t]
	\centering
	\includegraphics[width=\linewidth]{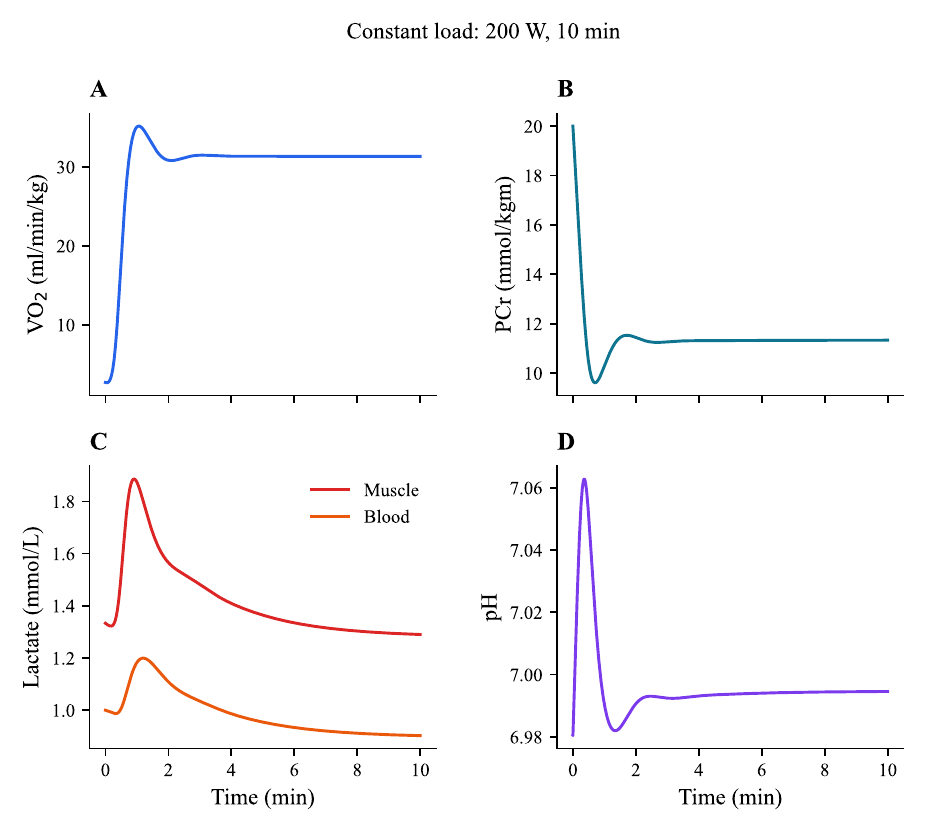}
	\caption{Dynamic simulation of constant-load exercise (\qty{200}{\watt}, \qty{10}{\minute}).
	For the athlete parameters used here (\qty{75}{\kilo\gram}, \qty{30}{\percent}~\acs{amm}, $\vomax = \qty{60}{\milli\litre\per\minute\per\kilo\gram}$, $\vlamax = \qty{0.7}{\milli\mole\per\litre\per\second}$), \qty{200}{\watt} is sub-\acs{mlss}, so all four traces converge to a steady state.
	(A)~Oxygen uptake with mono-exponential on-kinetics.
	(B)~Phosphocreatine buffering the initial ATP deficit.
	(C)~Muscle (red) and blood (orange) lactate reaching steady state.
	(D)~Intracellular pH response.
	Supra-\acs{mlss} behaviour (continuous lactate accumulation) is shown separately in Figs.~\ref{fig:steptest} and \ref{fig:sprint}.}
	\label{fig:dynamic}
\end{figure}

At constant-load exercise onset, $\vo$ rises mono-exponentially toward the Michaelis--Menten steady-state target (Fig.~\ref{fig:dynamic}A).
PCr drops rapidly, buffering ATP demand before oxidative phosphorylation reaches steady state (Fig.~\ref{fig:dynamic}B).
Muscle and blood lactate rise transiently and stabilise at the sub-\ac{mlss} steady-state level (Fig.~\ref{fig:dynamic}C; the chosen \qty{200}{\watt} are sub-\ac{mlss} for the athlete parameters used). pH falls with lactate accumulation, partially inhibiting glycolysis (Fig.~\ref{fig:dynamic}D).
Supra-\ac{mlss} behaviour (continuous lactate accumulation without steady state) is shown in Figs.~\ref{fig:steptest} and \ref{fig:running}.

\subsection{MLSS identification}

\begin{figure}[t]
	\centering
	\includegraphics[width=\linewidth]{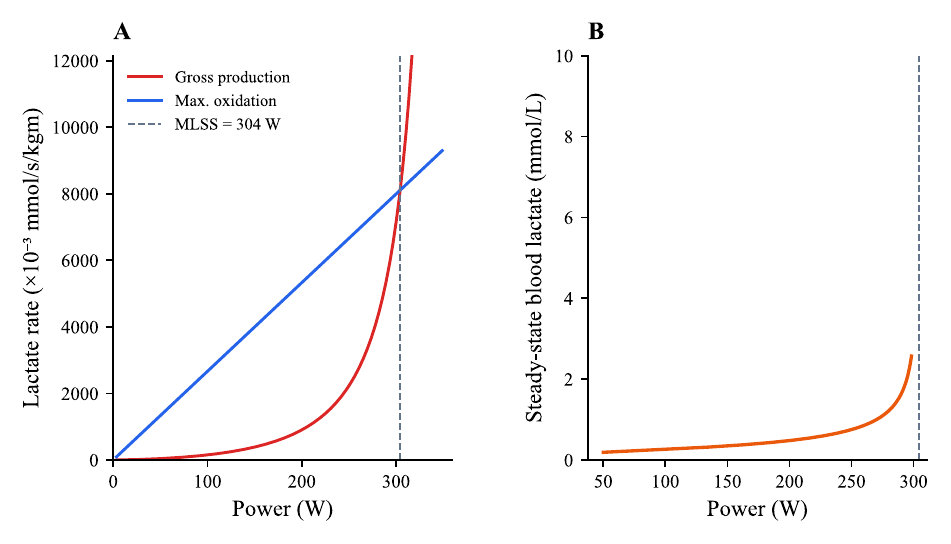}
	\caption{Steady-state model output.
	(A)~Gross lactate production (red) and maximal oxidation capacity (blue) as functions of power; their intersection defines the \acs{mlss} (dashed).
	(B)~Steady-state blood lactate concentration diverging as power approaches the \acs{mlss} in the one-compartment formulation.}
	\label{fig:steady}
\end{figure}

The one-compartment steady-state model identifies the \ac{mlss} as the crossing point of gross lactate production and maximal oxidation capacity (Fig.~\ref{fig:steady}A; cf. Fig.~\ref{fig:maxlass}).
Below the \ac{mlss}, blood lactate rises steeply with intensity (Fig.~\ref{fig:steady}B).
The two-compartment model yields \ac{mlss} estimates within \qtyrange{5}{10}{\watt} of the one-compartment value, confirming congruence between the formulations.

\subsection{Dynamic vs steady-state comparison}

Below the \ac{mlss}, the dynamic simulation converges to the blood lactate predicted by the algebraic model within \qtyrange{3}{10}{\minute} of simulated time.
Above the \ac{mlss}, the steady-state model predicts no stable equilibrium, consistent with the continuous lactate rise in the dynamic simulation.
The steady-state model is approximately 5--10$\times$ faster than the dynamic model and is therefore preferable for diagnostic applications where transient behaviour is not of interest.

\subsection{Parameter sensitivity}

\begin{figure}[t]
	\centering
	\includegraphics[width=\linewidth]{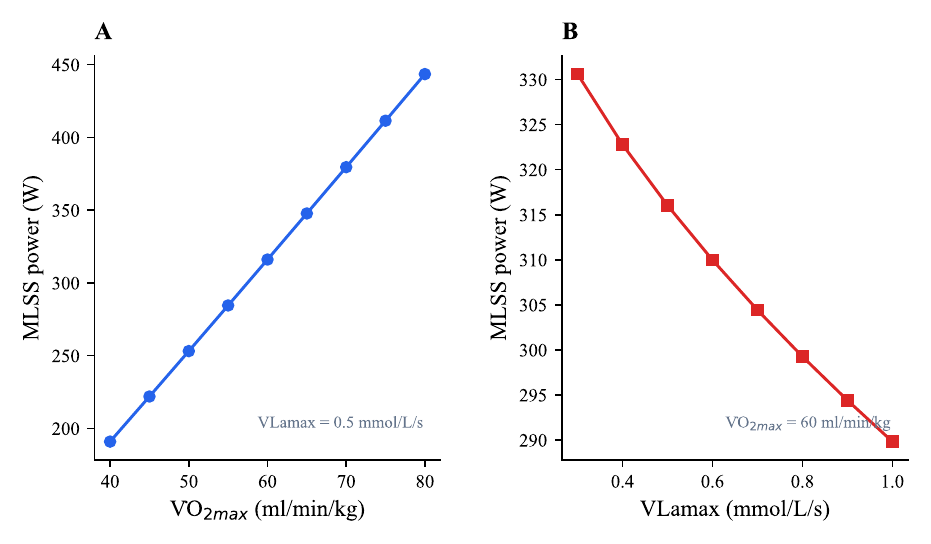}
	\caption{Parameter sensitivity of \acs{mlss} power.
	(A)~Effect of $\vomax$ at fixed $\vlamax = \qty{0.5}{\milli\mole\per\litre\per\second}$.
	(B)~Effect of $\vlamax$ at fixed $\vomax = \qty{60}{\milli\litre\per\minute\per\kilo\gram}$.
	Body mass $= \qty{75}{\kilo\gram}$, \qty{30}{\percent}~\acs{amm}.}
	\label{fig:sensitivity}
\end{figure}

Increasing $\vomax$ (\qtyrange{40}{80}{\milli\litre\per\minute\per\kilo\gram} in \qty{5}{\milli\litre\per\minute\per\kilo\gram} steps) at fixed $\vlamax = \qty{0.5}{\milli\mole\per\litre\per\second}$ shifts the \ac{mlss} power rightward approximately linearly (Fig.~\ref{fig:sensitivity}A).
Increasing $\vlamax$ (\qtyrange{0.3}{0.9}{\milli\mole\per\litre\per\second} in \qty{0.1}{\milli\mole\per\litre\per\second} steps) at fixed $\vomax = \qty{60}{\milli\litre\per\minute\per\kilo\gram}$ shifts the \ac{mlss} leftward and raises steady-state lactate (Fig.~\ref{fig:sensitivity}B), demonstrating the antagonistic influence of the two capacity parameters.

\subsection{Example simulations: step test}

\begin{figure}[t]
	\centering
	\includegraphics[width=\linewidth]{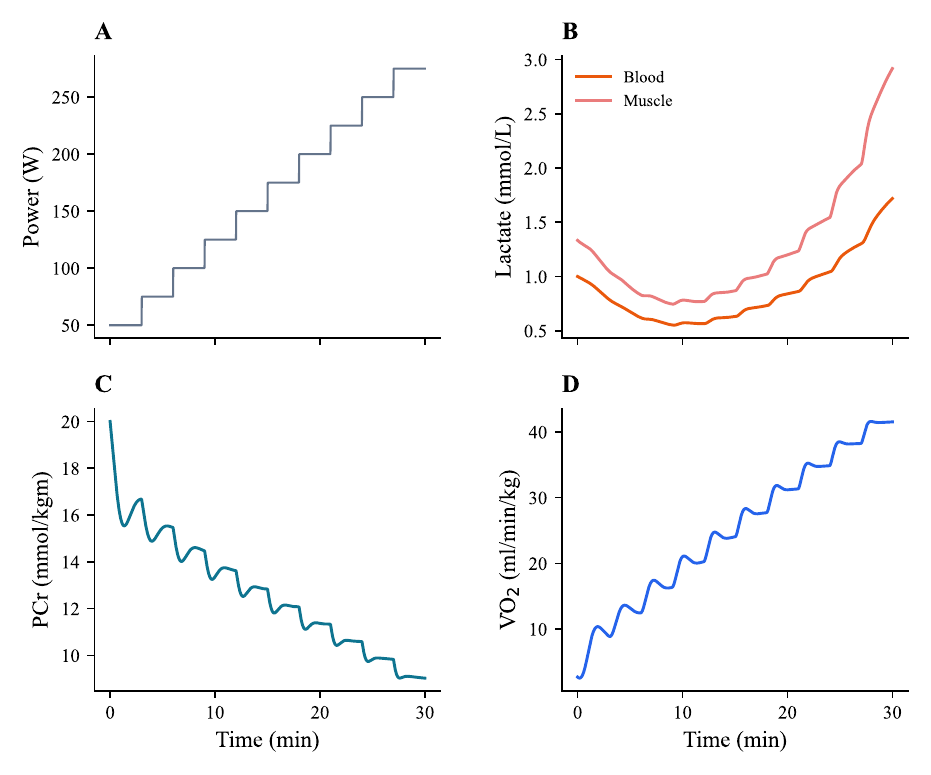}
	\caption{Simulated step test.
	(A)~Power protocol.
	(B)~Blood (orange) and muscle (red) lactate accumulation with the characteristic exponential lactate--power response.
	(C)~Phosphocreatine (teal, left axis) and $\vo$ (blue, right axis) across steps.}
	\label{fig:steptest}
\end{figure}

A simulated step test (\qty{50}{\watt} start, $+\qty{25}{\watt}$/step, \qty{180}{\second}/step, 10 steps) demonstrates progressive lactate accumulation and the transition from submaximal steady states to supra-\ac{mlss} non-equilibrium (Fig.~\ref{fig:steptest}).

\subsection{Example simulations: sprint and recovery}

\begin{figure}[t]
	\centering
	\includegraphics[width=\linewidth]{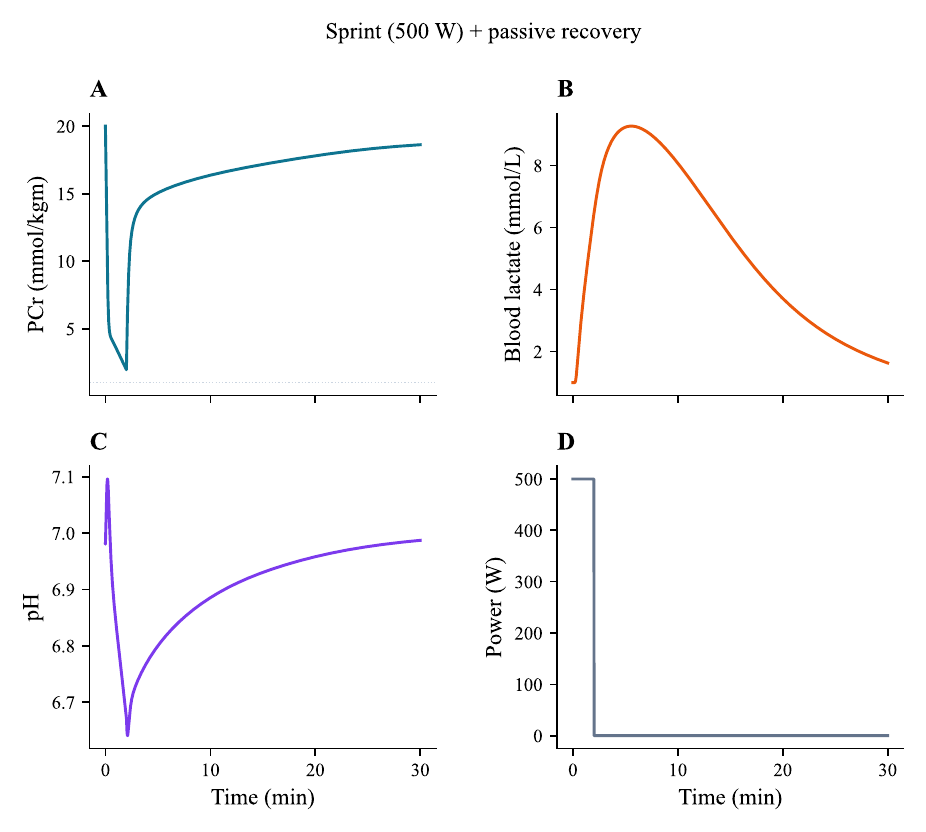}
	\caption{Sprint (\qty{500}{\watt}) with passive recovery.
	(A)~PCr depletion and exponential recovery.
	(B)~Blood lactate with post-exercise overshoot.
	(C)~pH nadir at peak lactate.
	(D)~Power protocol.}
	\label{fig:sprint}
\end{figure}

A \qty{500}{\watt} sprint until exhaustion followed by passive recovery shows PCr depletion to the exhaustion threshold, peak blood lactate of $\sim\qtyrange{12}{18}{\milli\mole\per\litre}$ at \qtyrange{2}{4}{\minute} post-cessation, and the subsequent exponential recovery of PCr and decline of lactate (Fig.~\ref{fig:sprint}).

\subsection{Plausibility across modes: running lactate dynamics}

\begin{figure}[t]
	\centering
	\includegraphics[width=\linewidth]{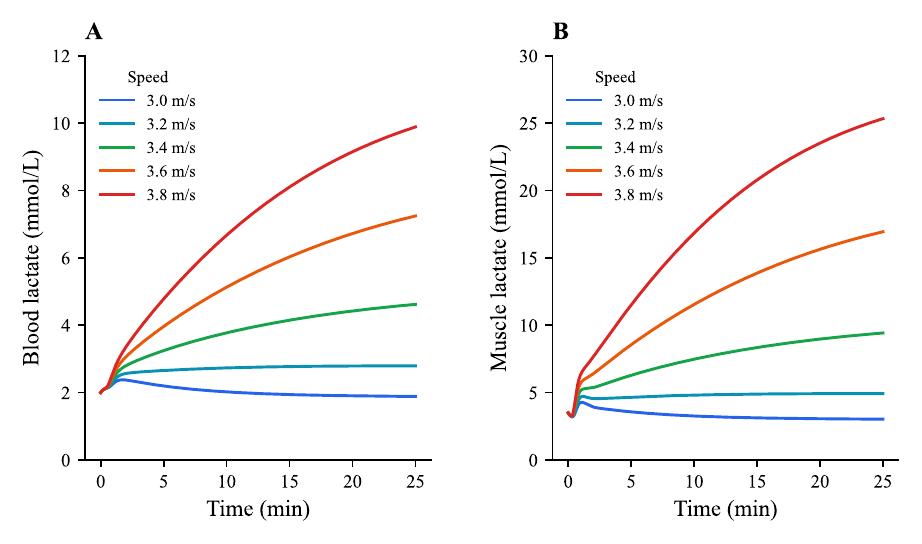}
	\caption{Running lactate dynamics at five constant velocities (\qtyrange{3.0}{3.8}{\metre\per\second}, \qty{25}{\minute}).
	(A)~Blood lactate: sub-\acs{mlss} velocities reach a steady state; supra-\acs{mlss} velocities show continuous accumulation.
	(B)~Muscle lactate.
	$\vo(v) = -1.1 + 12.4\,v$~\unit{\milli\litre\per\minute\per\kilo\gram} \parencite{maderheck1994}.
	Athlete: \qty{75}{\kilo\gram}, $\vomax = \qty{50}{\milli\litre\per\minute\per\kilo\gram}$, $\vlamax = \qty{0.7}{\milli\mole\per\litre\per\second}$.}
	\label{fig:running}
\end{figure}

To assess plausibility beyond cycling, the dynamic model was applied to running at five constant velocities (\qtyrange{3.0}{3.8}{\metre\per\second}, \qty{25}{\minute} each) using the running-specific $\vo$--velocity relationship of \textcite{maderheck1994}.
Below approximately \qty{3.2}{\metre\per\second}, blood lactate stabilises after a transient rise, consistent with a sustainable steady state.
Above this velocity, blood lactate increases continuously throughout the exercise bout, indicating supra-\ac{mlss} conditions (Fig.~\ref{fig:running}).
This qualitative separation of sub- and supra-\ac{mlss} responses reproduces the classical observation of \textcite{heck1985} in running and demonstrates that the model generates physiologically consistent output across exercise modes without re-parametrisation of the core metabolic equations.

\subsection{Self-consistency check: parameter recovery from synthetic step-test data}

\begin{figure}[htb]
	\centering
	\includegraphics[width=\linewidth]{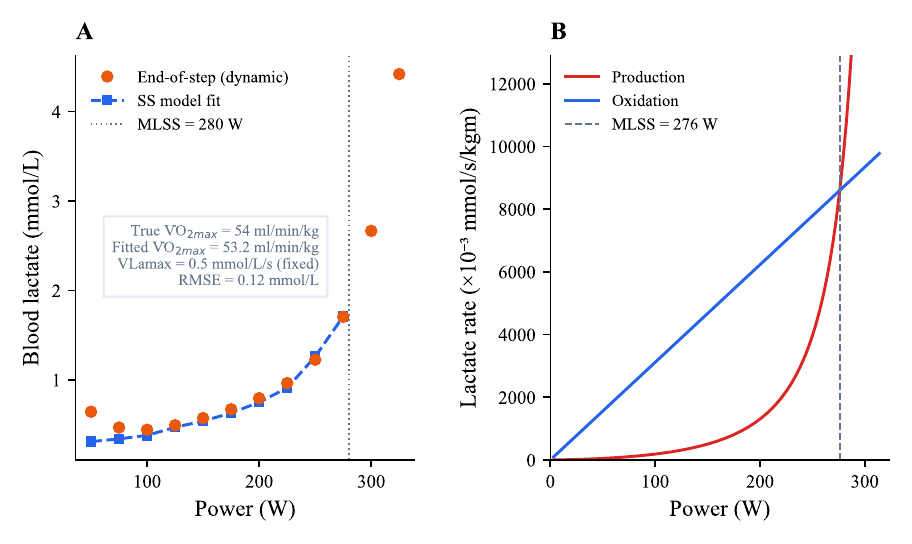}
	\caption{Self-consistency check: parameter recovery from synthetic step-test data.
	(A)~End-of-step blood lactate from the dynamic simulation (circles) and steady-state model fit (squares); dotted line: predicted \acs{mlss}.
	(B)~Fitted steady-state lactate production (red) and oxidation capacity (blue); their intersection defines the \acs{mlss} (\qty{282}{\watt}).
	Known input parameters: $\vomax = \qty{60}{\milli\litre\per\minute\per\kilo\gram}$, $\vlamax = \qty{0.7}{\milli\mole\per\litre\per\second}$ (fixed).
	Fitted $\vomax = \qty{58.4}{\milli\litre\per\minute\per\kilo\gram}$, RMSE $= \qty{0.11}{\milli\mole\per\litre}$.}
	\label{fig:recovery}
\end{figure}

As an internal consistency check on the steady-state solver, end-of-step blood lactate values were extracted from a graded exercise test (\qtyrange{50}{325}{\watt}, \qty{25}{\watt} increments, \qty{300}{\second} per step) \emph{generated by the dynamic model itself} with known parameters ($\vomax = \qty{60}{\milli\litre\per\minute\per\kilo\gram}$, $\vlamax = \qty{0.7}{\milli\mole\per\litre\per\second}$, body mass \qty{75}{\kilo\gram}, \qty{30}{\percent} active muscle mass; protocol scenario also used by Mader et al., in preparation).
$\vomax$ was then fitted to these synthetic data by minimising the sum of squared residuals against the two-compartment steady-state model, with $\vlamax$ fixed at the known value.
The fit recovered $\vomax = \qty{58.4}{\milli\litre\per\minute\per\kilo\gram}$ (true: \num{60.0}; RMSE $= \qty{0.11}{\milli\mole\per\litre}$; Fig.~\ref{fig:recovery}A).
The systematic underestimation reflects the fact that 5-minute stages do not fully reach equilibrium at higher intensities.
From the fitted parameters, the full steady-state lactate--power curve and the model-derived \ac{mlss} power (\qty{282}{\watt}) are obtained directly (Fig.~\ref{fig:recovery}B).

This procedure demonstrates internal consistency between the dynamic and steady-state solvers and confirms that the parameter-estimation machinery operates as intended.
It does \emph{not} demonstrate accuracy against experimental data: both the input data and the model are produced from the same equations.
The same workflow---sprint test for $\vlamax$, step test for $\vomax$, model-derived \ac{mlss}---is, however, the standard diagnostic application of the Mader model in applied sport science \parencite{heck2022}, and its implementation in \MetaboliSim{} is a transparent and reproducible alternative to manual calculation.

\section{Discussion}

\subsection{Computational implementation and reproducibility}

Although Mader and Heck computed individual model scenarios numerically from the 1980s onward, the underlying code was never released, and a commercial tool (INSCYD) covers only the steady-state formulation without disclosing its numerical methods.
\MetaboliSim{} is the first open implementation encompassing both formulations.
Python was chosen for its wide adoption in scientific computing, its numerical ecosystem (NumPy, SciPy, Pandas) and its accessibility to researchers without a computer-science background.

The frozen-coefficient \ac{rk4} follows the computational strategy described by \textcite{mader2003}, in which all enzymatic rates are evaluated once per time step.
The system contains processes on disparate time scales (milliseconds for \ac{ck}\,/\,\ac{ak} equilibration vs. minutes for lactate diffusion) and may therefore exhibit stiffness; an empirical stiffness analysis (e.g. via the spectral radius of the Jacobian over the operating range) has not been performed in this work.
The frozen-coefficient strategy mitigates fast--slow coupling pragmatically; a future extension to a fully implicit stiff solver (\acs{bdf}, Radau) is straightforward through SciPy and would be the cleaner numerical treatment.
The adaptive \ac{rkf45} method is already included for users requiring guaranteed error control.

Immutable parameter dataclasses, deterministic integration and automated test suites ensure that any simulation is exactly reproducible from its inputs.
The modular architecture (Fig.~\ref{fig:architecture}) means that the rate equations can be verified against published formulae without running the integrator, that the integrator can be tested independently, and that the user interface can be replaced without touching the model code.
Note that the per-simulation execution times reported in Results (\qtyrange{50}{150}{\milli\second}) refer to the core model evaluation only; end-to-end latency in the interactive web interface is dominated by Plotly rendering, reactive re-computation in the Shiny framework, and network round-trip rather than by the model itself, and is therefore noticeably higher.

\subsection{Physiological interpretation}

The model reproduces the core regulatory dynamics described by \textcite{mader2003} and \textcite{heck2022}.
The \ac{adp}-mediated coupling between demand and supply gives rise to $\vo$ on-kinetics, intensity-dependent lactate accumulation, \ac{mlss} emergence and glycogen-dependent threshold shifts.
The pH feedback on glycolysis limits the maximal glycolytic rate under acidosis, creating the self-limiting character of supra-\ac{mlss} exercise.
Whether the specific Hill exponents and half-inhibition constants accurately represent in vivo \ac{pfk} regulation remains open; these are effective parameters tuned to system-level behaviour rather than direct enzyme measurements.

The two formulations complement each other: the steady-state model is efficient for diagnostics and parameter fitting, and its closed-form structure makes the dependence on the underlying kinetic parameters transparent (cf. the $\maxLass$ section).
The dynamic model captures transient phenomena inaccessible to the algebraic formulation.

\subsection{Scope and limitations}

The Mader model treats the working muscle as a single homogeneous compartment and does not account for fibre-type heterogeneity, regional perfusion differences, or contributions of non-exercising tissues to whole-body lactate metabolism.
The power--ATP coupling assumes constant mechanical efficiency, which is an approximation at extreme power outputs.

Many kinetic constants are effective values derived from model fitting rather than independent measurements.
Changes in these parameters can substantially alter predictions near the \ac{mlss}.
The two-parameter characterisation ($\vomax$, $\vlamax$) achieves practical utility but can yield non-unique solutions when fitting both simultaneously to sparse data.

The present work establishes numerical correctness and qualitative physiological plausibility.
It does \emph{not} establish quantitative validity.
The parameter-recovery example (Fig.~\ref{fig:recovery}) is an internal consistency check on the solver, not a validation against measured athletes: both the input data and the model that generated them rest on the same equations.
A formal comparison against experimental data---measured $\vo$ kinetics, invasively determined muscle metabolites, or direct \ac{mlss} determination in athletes---has not been performed in this work and remains a primary objective for future use of the open implementation.
Without such external validation, the quantitative accuracy of specific simulation outputs should be interpreted with appropriate caution.

The model was developed primarily for cycling.
Extension to other modes (running, rowing, swimming) requires mode-specific efficiency relationships and potentially additional physiological mechanisms; the running example in Fig.~\ref{fig:running} uses the $\vo$--velocity relationship of \textcite{maderheck1994} and demonstrates that the core metabolic equations do not require re-parametrisation.

\subsection{What \texorpdfstring{\MetaboliSim}{MetaboliSim} enables}

The principal contribution of this work is not a new model but a transparent computational infrastructure for an established one.
By making the complete Mader model available as documented, tested and openly licensed code, \MetaboliSim{} removes the principal obstacle to external validation studies: independent groups can now compare model predictions to their own experimental data without re-implementing the model from primary literature.
The same infrastructure supports systematic parameter sensitivity analysis (Fig.~\ref{fig:sensitivity}), prospective simulation of complex exercise protocols (intervals, pacing strategies, repeated sprints), and use as a teaching tool through the interactive web interface.

\section{Conclusions}

\MetaboliSim{} provides a transparent, reproducible and extensible Python implementation of the Mader model of muscular energy metabolism, encompassing both the dynamic \ac{ode} formulation and the algebraic steady-state formulation.
The software is numerically verified against published reference values and produces output consistent with qualitative exercise physiology.
It supports step-test analysis, \ac{mlss} estimation, parameter fitting, individualised simulation and scenario-based performance modelling.
By making the complete implementation available as documented, tested and openly licensed code, \MetaboliSim{} enables---for the first time---external experimental validation of the Mader framework by independent groups, and extension of the model by the sport science and computational physiology communities.

\section*{Declarations}

\paragraph{Funding.} No external funding was received for this work.

\paragraph{Conflicts of interest.} The authors declare no conflicts of interest.

\paragraph{Data availability.} The source code, default parameter sets, test suite and documentation are publicly available under the \acf{agpl} (AGPL-3.0-or-later) licence at \url{https://codeberg.org/3phos/metabolisim}.

\nocite{*}
\printbibliography

\newpage
\appendix

\section{Detailed rate equations}
\label{app:A}

This appendix lists the equations underlying the right-hand side $f$ in Eq.~\eqref{eq:M1}.
The derivation of each equation is given in \textcite{mader2003} and \textcite{heck2022}; the form below follows \textcite{heck2022} (Eqs.~4.1--4.16).
All concentrations are per kg active muscle mass ($\mathrm{kg_m}$) unless stated otherwise.

\subsection{Creatine-kinase and adenylate-kinase equilibria}

At each instant, ATP, ADP, AMP and inorganic phosphate ($\mathrm{P_i}$) are determined by two coupled equilibria subject to conservation constraints.
The \ac{ck} equilibrium relates the ATP/ADP ratio to the $\mathrm{PCr}/\mathrm{P_i}$ ratio through the Lohmann constant $M_1$ \parencite{veech1979}:
\begin{equation}
	\frac{[\mathrm{ATP}]}{[\mathrm{ADP}]} = Q = M_1 \cdot \frac{[\mathrm{PCr}]}{[\mathrm{P_i}]},
	\tag{A1}\label{eq:A1}
\end{equation}
where $M_1 = [\mathrm{H^+}] \cdot M_2$, $M_2 = \num{1.66e9}$, $[\mathrm{P_i}] = S_C - [\mathrm{PCr}]$ (creatine conservation, $S_C = \qty{23.0}{\milli\mole\per\kgm}$), and $Q$ is the auxiliary ATP/ADP ratio.
Combined with the adenylate-kinase equilibrium ($M_3 = [\mathrm{ATP}][\mathrm{AMP}]/[\mathrm{ADP}]^2 = 0.96$) and the adenine nucleotide conservation $[\mathrm{ATP}] + [\mathrm{ADP}] + [\mathrm{AMP}] = S_A = \qty{6.0}{\milli\mole\per\kgm}$:
\begin{equation}
	[\mathrm{ADP}] = \frac{S_A \cdot Q}{M_3 + Q + Q^2},
	\tag{A2}\label{eq:A2}
\end{equation}
\begin{equation}
	[\mathrm{ATP}] = Q \cdot [\mathrm{ADP}], \qquad
	[\mathrm{AMP}] = \frac{M_3}{Q} \cdot [\mathrm{ADP}].
	\tag{A3}\label{eq:A3}
\end{equation}
The free enthalpy of ATP hydrolysis is
\begin{equation}
	\Delta G_{\mathrm{ATP}} = \Delta G^\circ + R\,T \ln\!\left(
	10^3 \cdot M_1 \cdot \frac{[\mathrm{PCr}]}{[\mathrm{P_i}]^2} \right),
	\tag{A4}\label{eq:A4}
\end{equation}
with $\Delta G^\circ = \qty{+30500}{\joule\per\mole}$, $R = \qty{8.314}{\joule\per\mole\per\kelvin}$, $T = \qty{310}{\kelvin}$.

\subsection{Rate equations}

\paragraph{Oxidative ATP supply.}
Michaelis--Menten kinetics with Hill exponent~2 \parencite[Eq.~4.12]{heck2022}:
\begin{equation}
	\dot{V}\mathrm{O}_{2,\mathrm{ss}} = \frac{\vomax}{1 + K_{s1}/[\mathrm{ADP}]^2},
	\tag{A5}\label{eq:A5}
\end{equation}
with $K_{s1} = \num{1.225e-3}~(\unit{\milli\mole\per\kgm})^2$.
Actual $\vo$ approaches $\dot{V}\mathrm{O}_{2,\mathrm{ss}}$ with first-order kinetics (default $k_{\vo} = \qty{0.2}{\per\second}$, $\tau = \qty{5}{\second}$).
An optional intensity-dependent time constant \parencite{drescher2012} is
\begin{equation}
	\tau = 0.0023\,(\%\vomax)^2 - 0.6352\,(\%\vomax) + 51.196.
	\tag{A5a}\label{eq:A5a}
\end{equation}

\paragraph{Glycolytic ATP supply.}
\textcite[Eq.~4.16]{heck2022}:
\begin{equation}
	\nu_{\mathrm{La}} = \frac{\vlamax \cdot f_{\mathrm{gly}}}{
		\bigl(1 + [\mathrm{H^+}]^3/K_{s3}\bigr)\bigl(1 + K_{s2}/[\mathrm{ADP}]^3\bigr)}.
	\tag{A6}\label{eq:A6}
\end{equation}
\ac{adp} activation and pH inhibition both follow Hill kinetics with exponent~3.
The glycogen modulation factor reduces $\vlamax$ sigmoidally as glycogen is depleted: $f_{\mathrm{gly}} = 1/\bigl(1 + (K_\mathrm{gly} \cdot \mathrm{Gly}_\mathrm{full}/\mathrm{Gly})^3 \bigr)$, where $K_\mathrm{gly}$ is the activation constant indicating the relative storage level at which glycolysis is inhibited at half maximum, $\mathrm{Gly}_\mathrm{full}$ is the glycogen storage capacity (\SI{83.3}{\milli\mole\per\kgm}, equivalent to \SI{15}{\gram\per\kgm}, and $\mathrm{Gly}$ is the current muscle glycogen concentration in \si{\milli\mole\per\kgm})
Glycogen also modulates $\vomax$ through a fourth-root function bounded at $\sim\qty{20}{\percent}$ reduction at full depletion.

\paragraph{Intracellular pH.}
Non-bicarbonate buffer equilibrium \parencite{maderheck1994}:
\begin{equation}
	\mathrm{pH} = \mathrm{pH_{base}} + \frac{0.8\,[\mathrm{P_i}] - [\mathrm{La}]_{\mathrm{kgm}}}{\beta_{\mathrm{NB}}}
	- 0.55 \log_{10}(P_{\mathrm{CO_2}}),
	\tag{A7}\label{eq:A7}
\end{equation}
with $\mathrm{pH_{base}} = 7.85$ (dimensionless), $\beta_{\mathrm{NB}} = \qty{54.0}{\milli\mole}\,\mathrm{H^+}/(\mathrm{pH}\cdot\mathrm{kg_m})$, $[\mathrm{La}]_{\mathrm{kgm}} = [\mathrm{La}]_m \cdot V_{\mathrm{rel}}$, and $P_{\mathrm{CO_2}} = \min(40 + 55\,\vo/\vomax, 150)~\unit{\mmHg}$.

\paragraph{Lactate oxidation.}
Saturating \ac{pdh} kinetics:
\begin{equation}
	\nu_{\mathrm{La,ox}} = \frac{K_{\mathrm{LaO_2}} \cdot \dot{V}\mathrm{O}_{2,\mathrm{ml/min}}}{1 + K_{\mathrm{el,ox}}/[\mathrm{La}]_m^2},
	\tag{A8}\label{eq:A8}
\end{equation}
where $\dot{V}\mathrm{O}_{2,\mathrm{ml/min}}$ is the current oxygen uptake converted from ATP equivalents via the P/O quotient, $K_{\mathrm{LaO_2}} = \qty{0.01475}{\milli\mole}/ \mathrm{ml}\,\mathrm{O_2}$ and $K_{\mathrm{el,ox}} = 2.0~(\unit{\milli\mole\per\litre})^2$.
The total oxidation is split into a muscle fraction ($2/3$) and a blood fraction ($1/3$).

\paragraph{Gluconeogenesis.}
Inhibited by \ac{adp}, activated by lactate:
\begin{equation}
	\nu_{\mathrm{La,res}} = \frac{v_{\max}}{
		\bigl(1 + [\mathrm{ADP}]^2/K_{\mathrm{ADP1}}\bigr)\bigl(1 + K_{\mathrm{VLares}}/[\mathrm{La}]_{\mathrm{kgm}}^2\bigr)}.
	\tag{A9}\label{eq:A9}
\end{equation}
Each mmol of resynthesised lactate costs \num{3.0} mmol ATP.

\subsection{Power-to-ATP-demand coupling}

ATP demand is derived from a polynomial $\vo$--power relationship.
For cycling:
\begin{equation}
	\dot{V}\mathrm{O}_{2,\mathrm{load}} = \frac{c_0 + c_1 \cdot P}{m_{\mathrm{body}}},
	\tag{A10}\label{eq:A10}
\end{equation}
with $c_0 = \qty{250}{\milli\litre\per\minute}$ (baseline) and $c_1 = K_{s4} = \qty{11.7}{\milli\litre}\,\mathrm{O_2}/(\unit{\minute}\cdot\unit{\watt})$.
Conversion to ATP demand per unit active muscle mass: $b_{\vo} = \mathrm{P/O} \cdot 2/22.4 = \num{0.2321}~\mathrm{mmol\,ATP}/\mathrm{ml}\,\mathrm{O_2}$.
For running, $c_1$ represents the $\mathrm{O_2}$ cost per unit velocity.

\subsection{ODE system}

The five differential equations governing the dynamic model are:
\begin{align}
	\frac{d\mathrm{GP}}{dt} &= \vo \cdot b_{\vo} + \nu_{\mathrm{La}} \cdot b_{V\mathrm{La}}
	- \max(\dot{n}_{\mathrm{demand}} - P_{\mathrm{rest,body}}, 0)
	- P_{\mathrm{rest,muscle}} - 3\,\nu_{\mathrm{La,res}},
	\tag{A11}\label{eq:A11} \\[4pt]
	\frac{d\vo}{dt} &= k_{\vo}\,(\dot{V}\mathrm{O}_{2,\mathrm{ss}} - \vo),
	\tag{A12}\label{eq:A12} \\[4pt]
	\frac{d[\mathrm{La}]_m}{dt} &= \frac{\nu_{\mathrm{La}} - \nu_{\mathrm{La,ox,m}} - 0.6\,\nu_{\mathrm{La,res}}}{V_{\mathrm{rel}}}
	- K_1\,([\mathrm{La}]_m \cdot V_{\mathrm{rel}} - [\mathrm{La}]_b),
	\tag{A13}\label{eq:A13} \\[4pt]
	\frac{d[\mathrm{La}]_b}{dt} &= V^*_{\mathrm{rel}}\,K_1\,([\mathrm{La}]_m \cdot V_{\mathrm{rel}} - [\mathrm{La}]_b)
	- \nu_{\mathrm{La,ox,b}} - 0.4\,\nu_{\mathrm{La,res}}/V_{\mathrm{rel}},
	\tag{A14}\label{eq:A14} \\[4pt]
	\frac{d\mathrm{Gly}}{dt} &= \frac{-\nu_{\mathrm{La}}/(2\,b_{V\mathrm{La}}) + 0.5\,\nu_{\mathrm{La,res}}}{5.555}.
	\tag{A15}\label{eq:A15}
\end{align}
Diffusion: $K_1 = K_{\mathrm{dif}} \cdot [\mathrm{La}]_b^{-1.4}$ \parencite{mader2003}.
$V^*_{\mathrm{rel}} = \mathrm{AMM\%} / (\text{lactate space \%} - \mathrm{AMM\%})$.
The factor \num{5.555} converts mmol glycosyl units to grams.
The two resting metabolic terms $P_{\mathrm{rest,body}}$ and $P_{\mathrm{rest,muscle}}$ separate whole-body and active-muscle baseline turnover; the demand term is clamped to zero to prevent negative demand during recovery.

\subsection{Steady-state equations}

Setting all time derivatives in Eqs.~\eqref{eq:A11}--\eqref{eq:A15} to zero and solving the resulting algebraic system in the one-compartment limit yields
\begin{equation}
	\nu_{\mathrm{La,ss}} = \frac{\vlamax}{1 + K_{s2}\bigl((\vomax - \vo)/(K_{s1}\,\vo)\bigr)^{1.5}},
	\tag{A16}\label{eq:A16}
\end{equation}
\begin{equation}
	[\mathrm{La}]_{b,\mathrm{ss}} = \sqrt{\frac{K_{\mathrm{el}}\,\nu_{\mathrm{La,ss}}}{\mathrm{PD}}}
	\tag{A17}\label{eq:A17}
\end{equation}
(both equivalent to Eqs.~(12) and (13) in earlier presentations of the Mader model).
The two-compartment formulation retains the full rate equations \eqref{eq:A6}--\eqref{eq:A9} and solves the four-variable algebraic system obtained from $d\mathrm{GP}/dt = d[\mathrm{La}]_m/dt = d[\mathrm{La}]_b/dt = d\vo/dt = 0$ by nested bisection and fixed-point iteration:
\begin{equation}
	0 = \nu_{\mathrm{La,ss}}(P) - \nu_{\mathrm{La,ox,m,ss}}(P)
	- 0.6\,\nu_{\mathrm{La,res,ss}}(P)
	- K_1\,([\mathrm{La}]_{m,\mathrm{ss}} \cdot V_{\mathrm{rel}} - [\mathrm{La}]_{b,\mathrm{ss}}) \cdot V_{\mathrm{rel}}.
	\tag{A18}\label{eq:A18}
\end{equation}

\section{Default parameter set}
\label{app:B}

Default values of the kinetic and structural parameters used in the \texttt{MaderConstants} dataclass.
Athlete-specific parameters (body mass, active muscle mass percentage, $\vomax$, $\vlamax$) are held in the \texttt{AthleteProfile} dataclass and are not listed here.
The ``Literature range'' column lists independent measurements where available; parameters marked ``effective; no independent measurement'' are fitting parameters of the Mader formulation and have no direct experimental correlate.

\begin{table}[t]
	\caption{Default parameters of \MetaboliSim{} with independent literature ranges where available.
	References are given in short form; full citations in the reference list.}
	\label{tab:params}
	\centering
	\small
	\begin{tabular}{@{}lll@{}}
		\toprule
		\textbf{Parameter} & \textbf{Default} & \textbf{Unit} \\
		\midrule
		$S_A$ (adenine pool) & \num{6.0} & \unit{\milli\mole\per\kgm} \\
		$S_C$ (creatine pool) & \num{23.0} & \unit{\milli\mole\per\kgm} \\
		$M_2$ (CK equil.) & \num{1.66e9} & --- \\
		$M_3$ (AK equil.) & \num{0.96} & --- \\
		$K_{s1}$ ($\vo$ half-sat.) & \num{1.225e-3} & $(\unit{\milli\mole\per\kgm})^2$ \\
		$K_{s2}$ (glycolysis) & \num{3.375e-3} & $(\unit{\milli\mole\per\kgm})^3$ \\
		$K_{s3}$ (pH inhib.) & \num{6.31e-21} & $(\unit{\mole\per\litre})^3$ \\
		P/O quotient & \num{2.6} & --- \\
		$b_{V\mathrm{La}}$ (ATP/La) & \num{1.4} & --- \\
		$k_{\vo}$ & \num{0.2} & \unit{\per\second} \\
		$K_{\mathrm{dif}}$ (diffusion base) & \num{0.065} & \unit{\per\second} \\
		$K_{\mathrm{LaO_2}}$ & \num{0.01475} & $\unit{\milli\mole}/\mathrm{ml}\,\mathrm{O_2}$ \\
		$K_{\mathrm{el,ox}}$ (PDH) & \num{2.0} & $(\unit{\milli\mole\per\litre})^2$ \\
		$\beta_{\mathrm{NB}}$ (buffer) & \num{54.0} & $\unit{\milli\mole}/(\mathrm{pH}\cdot\mathrm{kg_m})$ \\
		$V_{\mathrm{rel}}$ (muscle water) & \num{0.75} & \unit{\litre\per\kgm} \\
		$K_{s4}$ ($\mathrm{O_2}$/watt) & \num{11.7} & \unit{\milli\litre\per\minute\per\watt} \\
		$b_{V\mathrm{Lares}}$ (gluc.) & \num{3.0} & --- \\
		\bottomrule
	\end{tabular}
\end{table}

Five of the seventeen parameters ($K_{s1}$, $K_{s2}$, $K_{s3}$, $K_{\mathrm{dif}}$, $K_{\mathrm{LaO_2}}$, $K_{\mathrm{el,ox}}$) are effective fitting parameters of the Mader formulation with no direct experimental correlate; sensitivity to these parameters is addressed in the Limitations section.
The remaining parameters all fall within their respective independent literature ranges, with two exceptions: the default $S_C$ is at the lower end of the published range (23 vs. \qtyrange{30}{40}{\milli\mole}/\unit{\kilo\gram} wet muscle), and the default $k_{\vo}$ corresponds to a faster muscle-level kinetics than the pulmonary phase-II values typically reported (\qty{5}{\second} vs. \qtyrange{20}{50}{\second})---consistent with the model describing intramuscular ATP demand rather than pulmonary gas exchange.

\end{document}